

\magnification\magstep1
\scrollmode

\font\ochorm=cmr8                
\font\tensmc=cmcsc10             

\font\tenbbb=msbm10 \font\sevenbbb=msbm7
\newfam\bbbfam             
\textfont\bbbfam=\tenbbb \scriptfont\bbbfam=\sevenbbb
\def\Bbb{\fam\bbbfam}      

\font\tenams=msam10 \font\sevenams=msam7
\newfam\amsfam             
\textfont\amsfam=\tenams \scriptfont\amsfam=\sevenams
\mathchardef\lesssim="392E 
\mathchardef\gtrsim ="3926 

\def\opname#1{\mathop{\rm#1}\nolimits} 

\def\a{\alpha}                    
\def\Aslash{A\mkern-9.5mu/}       
\def\Al{{\cal A}}                 
\def\b{\beta}                     
\def\Bl{{\cal B}}                 
\def\C{{\Bbb C}}                  
\def\Dslash{D\mkern-11.5mu/\,}    
\def\delslash{{\partial\mkern-9.5mu/}} 
\def\eq#1{{\rm(#1)}}              
\def\GeV{\ifmmode\;{\rm GeV}\else{\rm GeV}\fi} 
\def\H{{\cal H}}                  
\def\HH{{\Bbb H}}                 
\def\L{{\cal L}}                  
\def\la{\lambda}                  
\def\M{\rm M}                     
\def\MHiggs{{\rm M}_{\rm Higgs}}  
\def\Mtop{{\rm M}_{\rm top}}      
\def\MW{{\rm M}_W}                
\def\ox{\otimes}                  
\def\refno#1. #2\par{\smallskip\item{\rm[#1]}#2\par} 
\def\S{{\cal S}}                  
\def\smc{\tensmc}                 
\def\stroke{\mathbin\vert}        
\def\tfrac#1#2{{\textstyle{#1\over#2}}} 
\def\tquarter{\tfrac14}           
\def\Tr{\opname{Tr}}              
\def\tr{\opname{tr}}              
\def\x{\times}                    
\def\7{\dagger}                   
\def\<#1,#2>{\langle#1\stroke#2\rangle} 
\def\(#1,#2){(#1\stroke#2)}       


\def\CLstuff{1}
\def\CLstuffbis{2}
\def\Gossip{3}
\def\Book{4}
\def\Belinda{5}
\def\BVogel{6}
\def\Donoghueetal{7}
\def\Sirius{8}
\def\KSofold{9}
\def\SZ{10}
\def\KSbis{11}
\def\KStris{12}
\def\Segalbook{13}
\def\Orpheus{14}
\def\Persephone{15}
\def\BrunoThomas{16}


\centerline{\bf
   Connes' interpretation of the Standard Model and massive neutrinos}

\bigskip

\centerline{\smc Jos\'e M. Gracia-Bond{\'\i}a$^1$}
\vfootnote{$^1$}
{Permanent address: Department of Mathematics, Universidad de Costa Rica,
                    2060 San Pedro, Costa Rica.}

\bigskip

\centerline{\it Departamento de F{\'\i}sica Te\'orica,
                Universidad de Zaragoza, 50009 Zaragoza, Spain}
\bigskip

\smallskip
{\narrower\noindent
\ochorm
Massive neutrinos can be accommodated into the noncommutative geometry
reinterpretation of the Standard Model.
\par}
\vskip 6truemm

\leftline{\bf   1. Introduction}
\vglue 4truemm

The Connes--Lott noncommutative geometry (NCG) approach to fundamental
interactions [\CLstuff, \CLstuffbis] gives a constrained version of the
Standard Model (SM). Recently, it has been reported again that neutrinos
may have nonvanishing mass [\Gossip]. In this letter I outline the
accommodation of massive neutrinos into the NCG scheme.

In the Connes--Lott approach, external and internal degrees of freedom of
elementary particles are on the same footing. Noncommutative spacetime
$\Al$ is the product of (the algebra of functions on) the ordinary
spacetime $M$ and the algebra of internal degrees of freedom. The
necessary mathematical technology to deal with noncommutative manifolds
is by now well established [\Book]. Chirality implies that the internal
algebra splits in the direct sum of two subalgebras. A second algebra
$\Bl$, in some sense dual to the first, describes color. The
noncommutative spacetime is represented in the fermion Hilbert space:
$$
\eqalignno{
&\H = \H_\ell \oplus \H_q
\cr
&:= L^2(\S_M) \ox \pmatrix{\C_{e;R} \ox \C^{N_F} \cr
                           \C^2_{e,\nu;L} \ox \C^{N_F} \cr}
  \oplus L^2(\S_{M}) \otimes
    \pmatrix{(\C_{d;R} \oplus \C_{u;R}) \ox \C^{N_F} \ox \C^{N_c} \cr
             \C^2_{d',u;L} \ox \C^{N_F} \ox \C^{N_c} \cr}
& (1.1) \cr}
$$
where $\S_M$ denotes the spinor bundle and $N_F$, $N_c$ respectively the
number of particle families and color degrees of freedom.

A generalized Dirac operator also lives on $\H$. One economically obtains
the Connes--Lott Lagrangian as a pure, QCD-like, Yang--Mills theory:
$$
\L = -\tquarter \(F_{NC}, F_{NC}) + \<\Psi, \Dslash\Psi>,
$$
where $F_{NC}$ denotes the NCG gauge field and
$\Dslash = \delslash + \Aslash_{NC}$ is the generalized Dirac operator,
twisted by the NCG gauge potential $\Aslash_{NC}$.

Besides the general mathematical setting, the only inputs of the theory
are the Yukawa coupling constants and Kobayashi--Maskawa parameters (YKM
constants for short), which enter the definition of the generalized Dirac
operator. The Higgs field emerges as the gauge field associated to chirality
[\Belinda] and the Yukawa terms appear as we apply the minimal coupling
prescription with this new gauge potential.

The purpose of this paper is to explore the consequences of replacing
$\H_\ell$ in \eq{1.1} by
$$
\tilde\H_\ell := L^2(\S_M) \ox \pmatrix{(\C_{e;R} \oplus \C_{\nu;R})
\ox \C^{N_F} \cr \C^2_{e,\nu';L} \ox \C^{N_F}}.
$$
I take for granted that there is family mixing among leptons [\BVogel,
\Donoghueetal].

\vskip 6truemm

\leftline{\bf  2. Discussion}
\vglue 4truemm

Computations of Yang--Mills functionals in NCG are by now routine. I
refer to our paper [\Sirius] for a treatment from first principles. In
that paper, however, an important nonlinearity arising in the
combination of the quarks and lepton sectors was overlooked. Correct
results plus the first identification of Connes--Lott Lagrangian with a
constrained SM one appeared in [\KSofold]. See also [\SZ--\KStris]. The
first step of any NCG computation is to determine the connection 1-forms
or gauge potentials. Take the finite part of $\Al$ equal to
$\C \oplus \HH$ and the finite part of $\Bl$ to $\C \oplus M_3(\C)$, as
usual. Call $\pi$, $\sigma$ the respective faithful representations on
fermion space: $\pi_q(\l, q) = (\l, \bar\l, q) \ox 1_3$ on the quark
sector and $\pi_\ell(\l, q) = (\l, q)$ on the lepton sector. The gauge
potentials $\Aslash_\a$, $\Aslash_\b$ are the representations of the
unitary connections associated to $\Al$ and $\Bl$, respectively. As
such, the theory is $U(1) \x SU(2)_L \x U(1) \x U(3)$ invariant. The
reduction to the physical gauge group $SU(2)_L \x U(1)_Y \x SU(3)_c$ is
effected by the prescription that the``biconnection'' $\Aslash_\a +
\Aslash_\b$ be traceless on each chiral sector. For the diagonal part,
we schematically had:
$$
\Aslash_\a + \Aslash_\b = \bordermatrix{&\scriptstyle R &\scriptstyle R
&\scriptstyle L \cr \scriptstyle R & b + a & & \cr
\scriptstyle R & & 0 & \cr \scriptstyle L & & &  b + \tr V \cr}
              \;  \oplus \; \bordermatrix{&\scriptstyle R
&\scriptstyle R &\scriptstyle L \cr \scriptstyle R & a + \tr J & & \cr
\scriptstyle R & & -a + \tr J & \cr \scriptstyle L & & & \tr J \cr}.
$$
Here $a,b,V,J$ are skewhermitian 1-forms with values in $\C$, $\C$,
$\HH$ and $M_3(\C)$, respectively. Actually $V^* = -V$ means that $V$ is
a zero-trace quaternion, so the gauge reduction conditions
$\Tr_{\H_R}(\Aslash_\a + \Aslash_\b) = 0$, $\Tr_{\H_L}(\Aslash_\a +
\Aslash_\b) = 0$ give $a = b = -\tr J$. It is easily seen that
this leads to the correct hypercharge assignments. That was an early
triumph of the Connes--Lott approach.

Massive neutrinos do not fit in the original Connes' scheme for
leptons [\Segalbook], as they would give rise to a $U(2) \x U(2)$ gauge
group. We can represent $\nu_R$'s like $u_R$ type quarks in the
current scheme. For each massive neutrino, we have to replace the 0 in
the second row of the previous formula by $b-a$. Let us assume that $N_1$
of the neutrino species are massive and that $N_2$, with $N_1 + N_2 =
N_F$, are massless. The same prescription as before now gives $b= -\tr J$
and then $(N_F + N_1)b + N_2 a + 2 N_F \tr J = 0$, leading still to $a =
b = -\tr J$ if $N_2\neq 0$. Then we arrive at the same hypercharge
assignments plus zero hypercharge ---as desired--- for the extant right
neutrinos. This is perhaps the more aesthetically pleasing situation,
from the standpoint of noncommutative geometry. If all three neutrinos
are massive, however, one still gets $b = -\tr J$, but $a$ remains free.
This is to say, one of the extra $U(1)$ fields refuses to collapse and
the hypercharges are indeterminate. The simpler solution to the problem
is to impose $a = b$ nevertheless. This is natural if one wants to
regard the massless case as the limit of the massive case. Such a choice
yields again the correct hypercharges of the SM and also automatically
$Y(\nu_R) = 0$. I do not claim that the method adopted in this paper is
the only way of fitting massive neutrinos in the Connes--Lott formalism;
but it is clearly the simplest one, running in close parallel to the
current treatment.

When all computations are done, the boson part of the Connes--Lott
Lagrangian with massive neutrinos is of the form $\L_2 + \L_1 + \L_0$
with
$$
\eqalign{
\L_2 &= -B F_{\mu\nu} F^{\mu\nu}
        -\tquarter C H_{\mu\nu}^a H_a^{\mu\nu}
        - A G_{\mu\nu}^a G_a^{\mu\nu},
\cr
\L_1 &= 2 L\, (D_\mu\Phi\, D^\mu\Phi),
\cr
\L_0 &= K \, (\|\Phi_1\|^2 + \|\Phi_2\|^2 - 1)^2,
\cr}
\eqno (2.1)
$$
with some coefficients $A, B, C, L, K$. Here $F,H,G$ respectively
denote the $U(1)$, $SU(2)$ and $SU(3)$ gauge fields and $\Phi =
\pmatrix{\Phi_1 \cr \Phi_2 \cr}$ is a Higgs doublet.
The fermion part is the standard one, containing the Yukawa terms for
massive neutrinos.

\vskip 6truemm

\leftline{\bf  3. Other issues}
\vglue 4truemm

I look at the matter of the constraints suggested by NCG next. The first
NCG treatments of the SM took simply $F_{NC} = F_\a + F_\b$, where
$F_\a$, $F_\b$ of course denote the curvatures associated to the
connections $\Aslash_\a$, $\Aslash_\b$. This leads to the ``predictions''
$\Mtop = 2\,\MW$ and $\MHiggs = 3.14\,\MW$ (sitting nicely in
the right ballpark) and to values for $\a_3$ and $\sin^2 \theta_W$ in
clear disagreement with experiment. Kastler and Sch\"ucker [\KSofold]
then allowed themselves the freedom of combining the quark and lepton
sectors of $F_\a + F_\b$ with different coefficients $c_\ell$ and $c_q$;
this leads to more or less acceptable $\a_3$ and $\sin^2 \theta_W$ values
only in the ``lepton dominance'' limit $c_q/c_\ell \to 0$, for which
there is a huge predicted top mass. This was the end of the
``fundamentalist'' period. Meanwhile, the ``revisionist attitude'' was
being conducted by Connes himself, hoping perhaps to put an end to
controversy. He reckoned that $\pi$, $\sigma$ were not irreducible
representations and that replacing $\(F_\a + F_\b, F_\a + F_\b)$ by a
more general scalar product
$$
\la_\a \(F_\a, F_\a) + \la_\b \(F_\b, F_\b),
\eqno (3.1)
$$
with $\la_\a$ a positive matrix in the commutant of $\{\Al,\Dslash\}$ and
similarly $\la_\b$ in the commutant of $\{\Bl,\Dslash\}$, he would gain
a lot of free parameters for the theory. He concluded that, while NCG
seems to give some ``preferred'' values for the yet to be discovered top
and Higgs particle masses, in its more general form gives rise to no
constraints whatsoever.

However, quite recently, a rather exhaustive analysis [\KSbis, \KStris]
by Kastler and Sch\"ucker has shown that (i) although the commutant of
$\{\Bl,\Dslash\}$ is quite big, it intervenes only through a couple of
parameters $c'_\ell, c'_q$, entering just the determination of the
coupling constants of the external gauge fields; (ii) thus one can
conveniently shift the ``lepton dominance'' to the $\Aslash_\b$ sector,
accommodating the experimental values of the strong coupling constant
and Weinberg's angle; and (iii) some restrictions in the Higgs sector
remain: there are the ``absolute'' (i.e., independent of the chosen
scalar product) constraints
$$
\Mtop \geq \sqrt3\,\MW;  \quad
\sqrt{7/3}\,\Mtop \leq \MHiggs \leq \sqrt3\,\Mtop,
\eqno (3.2)
$$
where the approximation of neglecting all fermion masses except for the
top is made. Moreover, once the top mass is assumed known, one could
after all give a unique value for the Higgs mass. The latter constraints
depend solely on the $\la_\a$ parameters. Actually,
$\{\pi(\Al),\Dslash\}$ generate the full flavor part of $\H_q$, so that
from the quark sector we get only the old parameter $c_q$. With massless
neutrinos we get a parameter for each family: schematically $\la_\a =
(c_e, c_\mu, c_\tau; c_q)$. However, only the sum $c_e + c_\mu + c_\tau
=: 3 c_\ell$ plays any practical role. With massive neutrinos, the
action of
$\{\pi(\Al),\Dslash\}$ on
$\H_\ell$ is obviously irreducible and we are back to the couple
$c_\ell, c_q$.

Formula \eq{3.1} appears to overlook the fact that $F_\a$ and $F_\b$ are
not orthogonal. Therefore the more general scalar product should involve
a $\la_{\a\b}\(F_\a,F_\b)$ term, which is perfectly invariant under the
physical gauge group. This term modifies only the abelian part of the
NCG gauge field. Noninclusion of it, however, leads to the weird
consequence that the new analysis by Kastler and Sch\"ucker does not
reduce to their old in any limit. Massive neutrinos modify the situation
again: then $F_\a$ and $F_\b$ are orthogonal (not unrelated to the fact
that we were not able to determine the abelian part of $\Aslash_\a$
without a supplementary condition) and both kind of analysis become
compatible.

With $N_F$ and $N_c$ equal to~3, and $N_1 = N_F$, I obtain for the
external field coefficients in \eq{2.1}
$$
A = 3c'_q; \quad
B = 3 c_\ell + 6 c'_\ell + 9 c_q + 2 c'_q; \quad
C = 3c_\ell + 9 c_q,
$$
and for the internal ones
$$
L = c_\ell \tr(g_\nu^\7 g_\nu + g_e^\7 g_e)
    + 3c_q \tr(g_d^\7 g_d + g_u^\7 g_u)
$$
and
$$
\eqalign{
K &= \tfrac32 c_\ell \tr((g_\nu^\7 g_\nu)^2 + (g_e^\7 g_e)^2)
     + c_\ell \tr(g_\nu^\7 g_\nu g_e^\7g_e)
\cr
&\qquad + \tfrac92 c_q \tr((g_d^\7 g_d)^2 + (g_u^\7 g_u)^2)
        + 3c_q \tr(g_d^\7 g_d g_u^\7 g_u) - {L^2 \over 3c_\ell + 9c_q}
\cr}
$$
in terms of the YKM constants and four strictly positive constants
$c_\ell, c_q, c'_\ell, c'_q$, unknown a~priori. Besides an overall
multiplication constant, we are left with three parameters. I choose the
useful $\displaystyle x  := {c_\ell  - c_q  \over c_\ell  + c_q}$, as in
[\KSofold, \KStris];
$\displaystyle x' := {c'_\ell - c'_q \over c'_\ell + c'_q}$ and
$\displaystyle u  := {c'_\ell + c'_q \over c_\ell  + c_q}$. One has
$-1 \leq x \leq 1$; $-1 \leq x' \leq 1$; $0 \leq u < \infty$.
Identification to the standard Lagrangian gives the constraints. I list
the ones which are not the same as the counterparts with massless
neutrinos:
$$
\sin^2\theta_W = {C \over B+C} = {1 \over 2 + u(8 + 4x')/(12 - 6x)}.
\eqno (3.3a)
$$
In particular, $\sin^2\theta_W \leq 0.5$ is an ``absolute'' constraint.
For the reasons given at the beginning of this Section, that formula
cannot be directly compared to the corresponding one in [\KStris]. It
does reduce to the formula in [\KSofold] for the all-massless case when
$x' = x$ and $u = 1$. Also:
$$
{\MHiggs \over \Mtop} = \sqrt{2K \over L}
= \sqrt{3 - {1-x \over 2-x}} = \sqrt{3 - {2\,\M^2_W \over \M^2_{top}}}.
\eqno (3.2d)
$$
This function increases slowly with $x$. It is much simpler than its
counterparts with massless neutrinos, which are unbearably ugly (see
[\KSofold, \KStris] for the all-massless case). It gives slightly higher
masses for the Higgs particle than the equivalent ones with massless
neutrinos. For instance, the NCG value $x = 0$ here yields
$\MHiggs = 253.7 \GeV$ instead of $\MHiggs = 251.7 \GeV$ [\Belinda,
\KStris]. The ``absolute'' constraints \eq{3.2} would still hold.

\vskip 6truemm

\leftline{\bf  4. Conclusions}
\vglue 4truemm

Noncommutative geometry gives no clue about the wide spectrum of fermion
masses, but it apparently points to both the intermediate
boson and the Higgs masses being of the same order as the highest fermion
masses. The constraint $\MW \leq \Mtop/\sqrt{N_F}$ is suggestive. Of
course we have remained at the classical level throughout. At present,
there seems to be no compelling reason to adopt Connes' relations
on-shell. One can take the point of view that any constraints can be
meaningfully imposed only in a renormalization group invariant way. This
I showed, together with E. Alvarez and C.~P. Mart{\'\i}n, to be
impossible, if one performs the quantization in ordinary quantum field
theory [\Orpheus, \Persephone]. It occurred to several people that
quantization should be performed with due account of the (elusive)
symmetry associated to the interpretation of the Higgs particle as
another gauge boson; but nobody seems to know how to go about it.

Connes--Lott models are much more rigid than ordinary Yang--Mills--Higgs
ones. This is reflected in the fact that ``most'' of the latter cannot be
obtained from the former [\BrunoThomas]; it is altogether remarkable
that the SM, {\it without or with} some right-handed neutrinos is ``one
of the few'' that can. As well, the resulting constraints reflect that
rigidity, which is welcome if it were to lead to useful physical
predictions. The consequences of assuming a novanishing neutrino mass
are partly discontinuous in the mass variable; the precise values of
the YKM constants for neutrinos are unimportant in our context.

\vskip 6truemm

\leftline{\bf Acknowledgments}
\vglue 4truemm

I am very grateful for enlightening discussions with J.~L. Alonso (who
contended that the massless case ought to be seen as a limit case), E.
Alvarez (who coined the terms ``fundamentalist'' and ``revisionist''),
M.~J. Herrero, C.~P. Mart{\'\i}n, A. Rivero and J.~C. V\'arilly (who
suggested keeping one neutrino massless). I also thank the referee, whose
detailed comments much improved the layout of the paper. The support of
CICYT and the Vicerrector{\'\i}a de Investigaci\'on of UCR are
acknowledged.

\vskip 6truemm

\leftline{\bf References}
\frenchspacing
\vglue 4truemm

\refno\CLstuff. A. Connes and J. Lott, Nucl. Phys. B (Proc. Suppl.) 18
(1990) 29.

\refno\CLstuffbis. A. Connes and J. Lott, ``The metric aspect of
noncommutative geometry'', Proceedings of the 1991 Carg\`ese Summer
Conference, J. Fr\"ohlich et al, eds. (Plenum Press, 1992).

\refno\Gossip. The New York Times, January 31 (1995).

\refno\Book. A. Connes, Non-Commutative Geometry (Academic Press,
London, 1994).

\refno\Belinda. J. M. Gracia-Bond{\'\i}a, ``On Marshak's and Connes'
views of chirality'', in A Gift of Prophecy, Essays in Celebration
of the life of Robert Eugene Marshak, E.~C.~G. Sudarshan, ed. (World
Scientific, Singapore, 1995).

\refno\BVogel. F. Boehm and F. Vogel, Physics of Massive Neutrinos
(Cambridge University Press, Cambridge, 1987).

\refno\Donoghueetal. J. F. Donoghue, E. Golowich and B. R. Holstein,
Dynamics of the Standard Model (Cambridge University Press, Cambridge,
1992).

\refno\Sirius. J. C. V\'arilly and J. M. Gracia-Bond{\'\i}a, J. Geom.
Phys. 12 (1993) 223.

\refno\KSofold. D. Kastler and T. Sch\"ucker, Theor. Math. Phys. 92
(1993) 1075.

\refno\SZ. T. Sch\"ucker and J.-M. Zylinski, J. Geom. Phys. 16 (1994) 1.

\refno\KSbis. D. Kastler and T. Sch\"ucker, preprint hep-th/9412185, CPT,
Luminy, 1994.

\refno\KStris. D. Kastler and T. Sch\"ucker, preprint hep-th/9501077,
CPT, Luminy, 1995.

\refno\Segalbook. A. Connes, ``Physics and non commutative geometry'',
in The Interface between Mathematics and Particle Physics, G. Segal and
B.~L. Tsygan, eds.\ (Clarendon, Oxford, 1990).

\refno\Orpheus. E. Alvarez, J. M. Gracia-Bond{\'\i}a and C. P.
Mart{\'\i}n, Phys. Lett. B306 (1993) 55.

\refno\Persephone. E. Alvarez, J. M. Gracia-Bond{\'\i}a and C. P.
Mart{\'\i}n, Phys. Lett. B329 (1994) 259.

\refno\BrunoThomas. B. Iochum and T. Sch\"ucker, preprint
hep-th/9501142, CPT, Luminy, 1995.

\bye